\documentclass[a4paper,10pt]{article}
\usepackage[utf8]{inputenc}
\usepackage{graphicx}
\usepackage{amssymb}
\usepackage{amsmath} 
\usepackage{bm}

\renewcommand{\vec}[1]{{\bm{#1}}}
\newcommand{\fr}[2]{{\displaystyle \frac{#1}{#2}}}

\newcommand{\pdiff}[2]{{\fr{\partial{#1}}{\partial{#2}}}}

\def\degr{\hbox{$^\circ$}}

\title{Features of the Flow Structure in the Vicinity of the Inner Lagrangian Point in Polars}
\author{P. B. Isakova\thanks{isakovapb@inasan.ru},
        A. G. Zhilkin, and
        D. V. Bisikalo\\
        \textit{\small Institute of Astronomy RAS, Moscow, Russia}}
\date{}
\begin{document}

\maketitle

\begin{abstract}

The structures of plasma flows in close binary systems whose accretors have strong intrinsic
magnetic fields are studied. A close binary system with the parameters of a typical polar is considered. The results of three dimensional numerical simulations of the material flow from the donor into the accretor Roche lobe are presented. Special attention is given to the flow structure in the vicinity of the inner Lagrangian point, where the accretion flow is formed. The interaction of the accretion flow material from the envelope of the donor  with the magnetic field of the accretor results in the formation of a hierarchical structure of the magnetosphere, because less dense areas of the accretion flow are captured by the magnetic field of the white dwarf earlier than more dense regions. Taking into account this kind of magnetosphere structure can affect analysis results and interpretation of the observations.

\end{abstract}

\section{Introduction}

Cataclysmic variables consist of a donor star (a low mass, late type star) and an accretor star (a
white dwarf) \cite{Warner:2003}. The donor’s envelope fills its Roche lobe. At the inner Lagrangian point, the pressure gradient is not balanced by the gravity. As a result, material starts to flow into the Roche lobe of the compact object. There is a wide class of cataclysmic variable in which the magnetic field can substantially influence the mass exchange and accretion processes. Magnetic cataclysmic variables can be divided into polars and intermediate polars. In intermediate polars, the magnetic field of the accretor is relatively weak ($10-1000~\text{kG}$), and an accretion disk can be formed in the system \cite{zbbUFN:2012}. In polars, the magnetic field of the accretor is strong ($> 1~\text{MG}$), and prevents the formation of the accretion disk. The material streaming from the donor forms a collimated stream that moves along the magnetic field lines and reaches one of the magnetic
poles of the accretor.

The magnetic fields in a close binary system does not only influence the motion of the stream from the donor envelope, but also controls the formation of this stream \cite{Norton:2004, Norton:2008, Campbell:1997}. This situation can arise in polars with strong magnetic fields (about $100~\text{MG}$ and higher), when the donor envelope is located partially in the magnetosphere of the white dwarf. The results of 3D numerical simulations of the flow structure in such systems \cite{zbbUFN:2012} show that the material streaming from the donor immediately splits into several flows, which move along magnetic field lines and reach the magnetic poles of the accretor. This flow pattern does not correspond to the classical picture of the formation of a stream from the donor to the Roche lobe of the accretor through the inner Lagrangian point \cite{Lubow:1975}. Taking into account the effect of the common envelope of a binary system can considerably change the flow pattern in the vicinity of the inner Lagrangian point \cite{Bisikalo:1997}. However, this effect is apparently more strongly manifest in intermediate polars, since almost all the material from the donor envelope falls on the accretor in polars.

In earlier studies, we have developed a 3D numerical model that can be used to study accretion in semidetached binary systems taking into account the magnetic field of the accretor [8]. In our previous
simulations, the material flow from the inner Lagrangian point towards the Roche lobe of the accretor
was determined by the boundary conditions. This model enables us to study only the influence of a
magnetic field on the already formed accretion stream. In the numerical model presented in our current study, the material flow from the donor is defined by a natural way, rather than by the imposed boundary conditions. Therefore, this model can be used to study the influence of the magnetic field on the formation of the accretion flow. 

The paper is organized as follows. In Section 2, we analyze the parameters that can affect the flow structure, such as the radius of the white dwarf magnetosphere. In Section 3, we discuss the formulation of the problem and describe our numerical model. The results of the numerical simulations are presented in Section 4. We discuss our main results in the Conclusion.

\section{Estimates of the Radius of the Accretor Magnetosphere}

A simple way to estimate the radius of the accretor magnetosphere in polar type systems is to
take it to be comparable to the corresponding Alfv{\`e}n radius $R_\text{m}$ \cite{Lipunov1987}. Let us consider the material stream moving from the vicinity of the Lagrangian point towards the accretor. This stream is stopped by the magnetic field at some distance from the accretor. We can estimate this distance by writing the condition for equality of the magnetic and dynamical pressures:
\begin{equation}\label{eq-Pm=Pd} 
 \fr{B_{*}^2}{8 \pi} = \rho v^2,
\end{equation}
where $v = \sqrt{2 G M_\text{a} / r}$ is the free-fall velocity, $G$  the gravitational constant, and $M_\text{a}$ the mass of the accretor. The density $\rho$ can be calculated from the mass transfer rate:  $\dot{M} = S \rho v$,  where $S$  is the cross section of the accretion stream. Generally, the magnetic field of the accretor in a typical polar can be described by a dipole field  \cite{Landau2}:
\begin{equation}\label{Bs}
 {\vec B}_{*} = \frac{\mu}{r^3} 
 \left[ 3 (\vec{d} \cdot \vec{n}) \vec{n} - \vec{d} \right],
\end{equation}
where $\mu = B_\text{a} R_\text{a}^3 / 2$  is the magnetic moment,  $B_\text{a}$  the magnetic field at the surface of the white dwarf,  $R_\text{a}$ the radius of the white dwarf, and ${\vec d}$  a unit vector determined by the axis of symmetry of the dipole field. The vector magnetic moment is ${\vec \mu} = \mu {\vec d}$ and the unit vector is  ${\vec n} = {\vec r} / r$. Let us assume that the accretor’s center is located at the coordinate origin. The components of the magnetic moment $\vec{\mu}$ are
\begin{equation}\label{eq-mu}
 \mu_x = \mu\sin\theta\cos\phi, \quad
 \mu_y = \mu\sin\theta\sin\phi, \quad 
 \mu_z = \mu\cos\theta, 
\end{equation}
where $\theta$  is the inclination of the vector $\vec{\mu}$  relative to the $z$, and $\phi$  is the angle between the projection of  $\vec{\mu}$  onto the equatorial plane of the binary system  ($xy$)  and the  $x$ axis. In our model, we assumed that the rotation and orbital motion of the components
is synchronous, so that the phase angle  $\phi$  is time independent. The magnetic field $\vec{B}_{*}$  is potential in the computational domain,  $\nabla \times \vec{B}_{*} = 0$, This enables us to partially exclude this field from the equations describing the structure of the magnetohydrodynamical
(MHD) flows \cite{mcb-book, Tanaka:1994, Powell:1999}.

The cross sectional area of the stream of material
from the donor in the vicinity of the inner Lagrangian
point can be estimated by the expression  \cite{Lubow:1975, mcb-book}:
\begin{equation}\label{eq-SL1} 
 S = \fr{\pi c_\text{s}^2}{4 \Omega^2} g_y\left( q \right) g_z\left( q \right),
\end{equation}
where $g_y\left( q \right)$ and $g_z\left( q \right)$ are functions that depend on the component mass ratio $q = M_\text{d}/M_\text{a}$ ($M_\text{d}$ is the mass of the donor) that are close to the unity and determine the size of the stream along the $y$ and $z$, axis, while $\Omega$ is the orbital angular velocity of the system. Using these relations and Eq. \eqref{eq-Pm=Pd}, we can obtain the estimate

\begin{equation}\label{eq-Rm} 
 R_\text{m} = \left[ \fr{B_\text{a}^4 R_\text{a}^{12} S^2}{128 \pi^2 G M_\text{a} \dot{M}^2} \right]^{1/11}.
\end{equation}
We estimated the magnetosphere radius using the parameters of a typical polar, taking as an example the orbital parameters of SS Cyg \cite{Giovannelli:1983}.  The donor (red dwarf) has the mass $M_\text{d} = 0.56~M_\odot$ and the effective temperature $T_\text{eff, d} = 4\,000~\text{K}$. The mass of the accretor is  $M_\text{a} = 0.97~M_\odot$,  its radius is $R_\text{a} = 7.3 \cdot 10^{-3}~R_\odot$, and its effective temperature is $T_\text{eff, a} = 37\,000~\text{K}$. The binary orbital period is  $P_\text{orb} = 2.4 \cdot 10^4~\text{s}$. The distance between the binary components is  $A = 2.05~R_\odot$, and the inner Lagrangian point is placed at a distance $r_{\text{L}_1} = 0.556~A$  from the accretor. The sound speed at the point  L$_1$  is defined to be  $c_\text{s} = 7.4 \cdot 10^5~\text{ cm/s}$, and the mass transfer rate through the vicinity of the inner Lagrangian point is $\dot{M} = 10^{-10}~M_\odot /~\text{yr}$.  Let us consider the case of a strong magnetic field, when the field at the surface of the white dwarf is $B_\text{a} = 10^8~\text{G}$.

These values give a cross section for the stream at the inner Lagrangian point of  $S = 0.000332~A^2$.  The corresponding diameter of the stream cross section is 2 $2\sqrt{S/\pi} = 0.0206~A$. The magnetosphere radius is $R_\text{m} = 0.14~A$. 

The material in the stream is stopped at the magnetosphere radius by the magnetic field. However,
it is clear that the magnetic field begins to influence the flow dynamics much earlier. We can estimate
the distance at which this influence begins using the expressions for the forces acting on the stream material in our model (see the next Section). There are two forces acting on the material in the Roche lobe of the accretor: the gravitational force of the white dwarf and the friction force from its magnetic field. Let us determine the distance $R_\text{w}$,  at which these forces become equal:
\begin{equation}\label{eq-Ffr=Fgr} 
 \fr{G M_\text{a}}{r^2} = \fr{v}{t_w}.
\end{equation}
Here, $t_w$  is the decay time scale for the transverse velocity, determined by the expression:
\begin{equation}\label{eq-tw} 
 t_w = \fr{4 \pi \rho}{B_{*}^2} \eta_w,
\end{equation}
$\eta_w$ is the magnetic viscosity coefficient associated with MHD wave turbulence,
\begin{equation}\label{eq-eta_w} 
 \eta_w = \alpha_w \fr{l_w B_{*}}{\sqrt{4 \pi \rho}},
\end{equation}
where $\alpha_w$ is a dimensionless coefficient, which we took to be $1/3$, corresponding to isotropic turbulence \cite{Landau8}, and $l_w = B_{*} / |\nabla B_{*}|$ is the characteristic spatial
scale of the wave pulsations. Given Eq. \eqref{Bs}  we can set $l_w = r / 3$. 
 Using the resulting relation, we find from \eqref{eq-Ffr=Fgr}
\begin{equation}\label{eq-Rw} 
 R_\text{w} = \left[ \fr{81 B_{*}^4 R_\text{a}^{12} S^2}{2 \pi^2 \alpha_w^4 G M_\text{a} \dot{M}^2} \right]^{1/11}.
\end{equation}

Note that this expression differs from \eqref{eq-Rm} only in the numerical coefficients. Substituting the parameter values gives $R_\text{w} = 0.46~A$, which is about a factor of three larger than the Alfv{\`e}n radius \eqref{eq-Rm}. It is easy to show that the friction force $f = v / t_w$ due to the
magnetic field of the accretor is the dominant force at the Alfv{\`e}n radius; its value relative to the gravitational force $g = G M_\text{a} / r^2$ is $f/g \left( R_\text{m} \right) = 25.5$. This means that the plasma flows inside this zone are fully governed by the magnetic field of the white dwarf.

\section{Formulation of the Problem}

We described the plasma flow structure in a magnetic close binary system using a non-inertial reference frame rotating about its center of mass with the orbital angular velocity $\Omega = 2 \pi /P_\text{orb}$,  where $P_\text{orb}$  is the orbital period. The force field acting on the material
in this frame is determined by the Roche potential, which has the form
\begin{equation}\label{eq-phi}
 \Phi = 
 -\frac{G M_\text{a}}{|\vec{r} - \vec{r}_\text{a}|}  
 -\frac{G M_\text{d}}{|\vec{r} - \vec{r}_\text{d}|}  
 -\frac{1}{2} 
 \left[ \vec{\Omega} \times 
 \left( \vec{r} - \vec{r}_\text{c} \right) 
 \right]^2,
\end{equation}
where the radius vectors $\vec{r}_\text{a}$, $\vec{r}_\text{d}$, and $\vec{r}_\text{c}$ define the
accretor’s center, the donor’s center, and the center of mass of the binary system, respectively. The first and second terms in \eqref{eq-phi}  describe the gravitational potential of the accretor and donor. The last term describes the centrifugal potential.

For the numerical simulations, we used a Cartesian coordinate system ($x$, $y$, $z$), with its origin coincident with the center of the accretor,  $\vec{r}_\text{a} = (0, 0, 0)$. The center of mass of the donor is placed at a distance  $A$ from the origin along the $x$ axis, $\vec{r}_\text{d} = (-A, 0, 0)$. The $z$ axis is along the orbital rotational axis of the system, such that the angular velocity vector has the components $\vec{\Omega} = (0, 0, \Omega)$.

The magnetic field of the white dwarf  $\vec{B}_{*}$ can be fairly strong in the magnetosphere region. Therefore, for convenience, we can represent the total magnetic field in the plasma $\vec{B}$ as a superposition of the background magnetic field  $\vec{B}_{*}$ and the field $\vec{b}$ induced by electric currents in the plasma, $\vec{B} = \vec{B}_{*} + \vec{b}$. In the finite-difference scheme, only the internal magnetic field of the plasma $\vec{b}$,  is calculated, enabling us to avoid an accumulation of errors from operations with large floating numbers in the numerical simulations.

The flow structure in magnetic close binary systems can be described by the set of equations \cite{zbbUFN:2012}:
\begin{equation}\label{eq-rho1}
 \pdiff{\rho}{t} + \nabla \cdot \left( \rho {\vec v} \right) = 0,
\end{equation}
\begin{equation}\label{eq-v1}
 \pdiff{{\vec v}}{t} + \left( {\vec v} \cdot \nabla \right) {\vec v} =
 -\frac{\nabla P}{\rho} -
 \frac{{\vec b} \times \nabla \times {\vec b}}{4 \pi \rho} -
 \nabla \Phi +
 2 \left( {\vec v} \times {\vec \Omega} \right) -
 \frac{\left( {\vec v} - {\vec v}_{*} \right)_{\perp}}{t_w},
\end{equation}
\begin{equation}\label{eq-b1}
 \pdiff{{\vec b}}{t} = \nabla \times \left[ {\vec v} \times {\vec b} +
 \left( {\vec v} - {\vec v}_{*} \right) \times {\vec B}_{*} -
 \eta\, \nabla \times {\vec b} \right],
\end{equation}
\begin{equation}\label{eq-e1}
 \rho \left[ \pdiff{\varepsilon}{t} +
 \left( {\vec v} \cdot \nabla \right) \varepsilon \right] =
 - P \nabla \cdot {\vec v} + 
 n^2 \left( \Gamma - \Lambda \right) +
 \frac{\eta}{4 \pi} \left( \nabla \times {\vec b} \right)^2,
\end{equation}
where $\rho$ is the density, ${\vec v}$ the velocity,  $P$  the pressure, $\varepsilon$  the internal energy per unit mass,  $n$  the number density, $\eta$ the coefficient of magnetic viscosity, and ${\vec v}_(*)$ the velocity of the magnetic field lines. The term  $2(\vec{v} \times \vec{\Omega})$  in the equation of motion \eqref{eq-v1} describes the Coriolis force. The density, energy, and pressure are related through the equation of state of an ideal gas:
\begin{equation}\label{eq-state}
 P = \left( \gamma - 1 \right) \rho \varepsilon,
\end{equation}
where $\gamma = 5/3$ is the adiabatic index. The energy equation ( \eqref{eq-e1}  takes into account the effects of radiative heating  $\Gamma$  and cooling $\Lambda$, as well as heating due to current dissipation  \cite{Cox1971, Dalgarno1972, Raymond1976, Spitzer1981}. Note that, in contrast to our previous papers  \cite{zb:2009, zbSMF:2010, zbMFG:2010, zbbUFN:2012},  we used the energy equation rather than the entropy equation \cite{Romanova2003, BisnovatyiKogan:2016} in this model.

Our model is based on the modified MHD approximation \cite{zbSMF:2010, zbbUFN:2012},  which is described in detail in our recent paper \cite{kzbUFN:2017}. This approximation corresponds
to the case of strong external magnetic fields, taking into account Alfv{\`e}n wave turbulence in the presence of small magnetic Reynolds numbers ($\text{R}_\text{m} \ll 1$) \cite{Braginsky1959}. In fact, plasma dynamics in strong external magnetic fields can be described by a relatively slow mean
motion of the particles along the magnetic field lines, a drift across the field lines, and the propagation of Alfv{\`e}n and magnetoacoustic waves with speeds that are very high compared to these background speeds Over a typical dynamical time scale, MHD waves can intersect the flow region multiple times. This makes it possible to investigate the mean flow pattern, considering the effect of rapid pulsations by analogy with wave MHD turbulence \cite{Zakharov:1965, Iroshnikov:1963, Kraichnan:1965}. To describe the slow proper motion of the plasma, it is necessary to distinguish the rapidly propagating fluctuations and apply a specific procedure for the ensemble averaging of the wave pulsations. We developed this model for MHD flows in application to polars and intermediate
polars in \cite{ zbmBYCam:2012, tilted1, tilted2, exhya1, tilted3, aeaqr1, exhya2}.

The last term in the equation of motion \eqref{eq-v1} describes the force acting on the plasma due to the accretor magnetic field. This affects only the plasma velocity perpendicular to the magnetic field lines \cite{Drell1965, Gurevich1978, Rafikov1999}.  The subscript $\perp$  denotes the velocity perpendicular to the magnetic field of the white dwarf $\vec{B}_{*}$. The vector $\vec{v}_{*}$ defines the velocity of the magnetic field lines due to the rotation of the white dwarf and the conductivity of the plasma. Since we have taken the rotation of the binary components to be synchronous, we can neglect the latter effect and set $\vec{v}_{*} = 0$.\footnote{ Estimates of the influence of the plasma conductivity on the field line velocity carried out in our recent
study \cite{exhya2} demonstrate that this effect is less than 20\% in the accretion disk. This effect should apparently be even less important in the accretion streams of polars.}  A strong external magnetic field acts like an effective fluid with which the plasma interacts, since the corresponding force is similar to the friction force between the components in a plasma consisting of several types of particles \cite{Frank-Kamenetsky1968, Chen1987}. Therefore, we can interpret this as a ''friction force'' between the plasma and magnetic field lines.

The following boundary and initial conditions were used in our model. In the donor envelope, we took
the velocity normal to the donor surface to be equal to the local sound speed, $v_{n} = c_\text{s}$  corresponding to an effective temperature for the donor of $4\,000~\text{K}$. The gas density in the donor envelope  $\rho(L_1)$ is determined by the mass transfer rate,
\begin{equation}\label{eq-Mdot}
 \dot{M} = \rho(L_1) v_{n} S,
\end{equation}
where the stream cross section is calculated using \eqref{eq-SL1}.

We imposed the following boundary conditions at the other boundaries of the computational domain. The density was set equal to $\rho_\text{b} = 10^{-8}~\rho(L_1)$, the temperature $T_\text{b}$  to the equilibrium temperature  $T_{*} = 11\,227~\text{K}$, and the magnetic field ${\vec B}_\text{b} = {\vec B}_{*}$. We specified a free outflow condition for the velocity ${\vec v}_\text{b}$ applying symmetrical boundary conditions when the velocity is directed outward, and specifying ${\vec v}_\text{b} = 0$ when the velocity is directed inward.

The accretor was taken to be a sphere of radius $0.0125~A$,  at whose boundary a free inflow boundary
condition was defined. The radius of the ''numerical'' star is about a factor of three larger than the
actual radius of the white dwarf  $R_\text{a}$, All the material that has intersected this boundary was taken to have fallen onto the white dwarf. The initial conditions in the computational domain were as follows: $\rho_0 = 10^{-8} \rho(L_1)$, $T_0 = T_{*}$, ${\vec v}_0 = 0$, and ${\vec B}_0 = {\vec B}_{*}$.

We carried out our computations using the Nurgush 2.0 3D parallel numerical code \cite{zb:2009, zbSMF:2010}\footnote{State registration number 2016663823.},  which is based on a Godunovtype finite-difference scheme with a high order approximation for the MHD equations. The solution was obtained in a computational domain with dimensions $-0.8~A \le x \le 0.8~A$, $-0.8~A \le y \le 0.8~A$, $-0.4~A \le z \le 0.4~A$ containing  $256 \times 256 \times 128$. cells. This computational domain fully encompasses the Roche lobe of the accretor, as well as part of the donor Roche lobe. This means that, in our model, the material flowing out from the donor in the vicinity of the inner Lagrangian point is determined by a natural manner, rather than by the  boundary conditions.

\section{Computational Results}

We will now present the results of our modeling of the flow structure for a typical polar (with a magnetic field $B_\text{a}=10^8~\text{G}$) 3D computations were carried out in the full computational domain, including the accretor. We also investigated the region near the inner Lagrange point in more detail. The computations were continued until the onset of the steady-state flow regime. The orientation of the axis of symmetry of the dipole magnetic field \eqref{eq-mu} was specified by the angles $\theta = 30\degr$, $\phi = 90\degr$.

\begin{figure}[ht!]  
\centering
\includegraphics[width=0.9\textwidth]{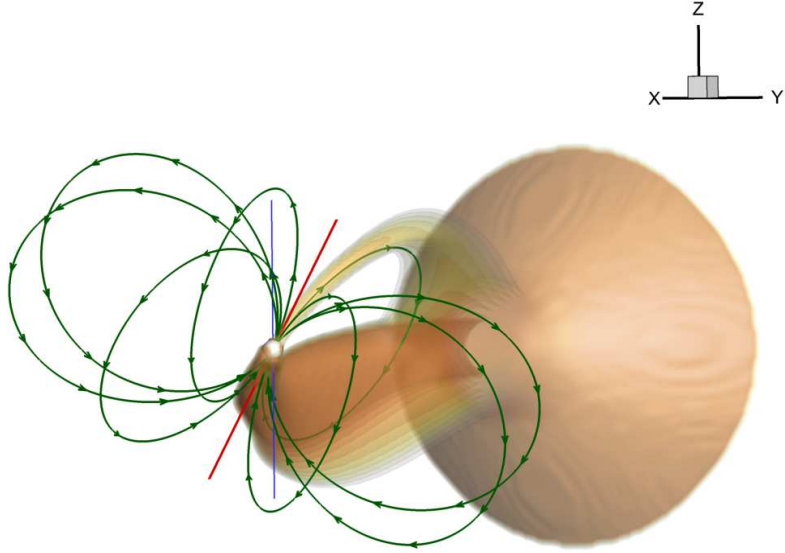}
\caption{The flow structure in a typical polar. Isosurfaces of logarithm density (color), magnetic field lines (lines with arrows), the magnetic axis (red line), and the rotation axis (blue line) are shown.}
\label{big}
\end{figure}

\begin{figure}[ht!]  
\centering
\includegraphics[width=0.9\textwidth]{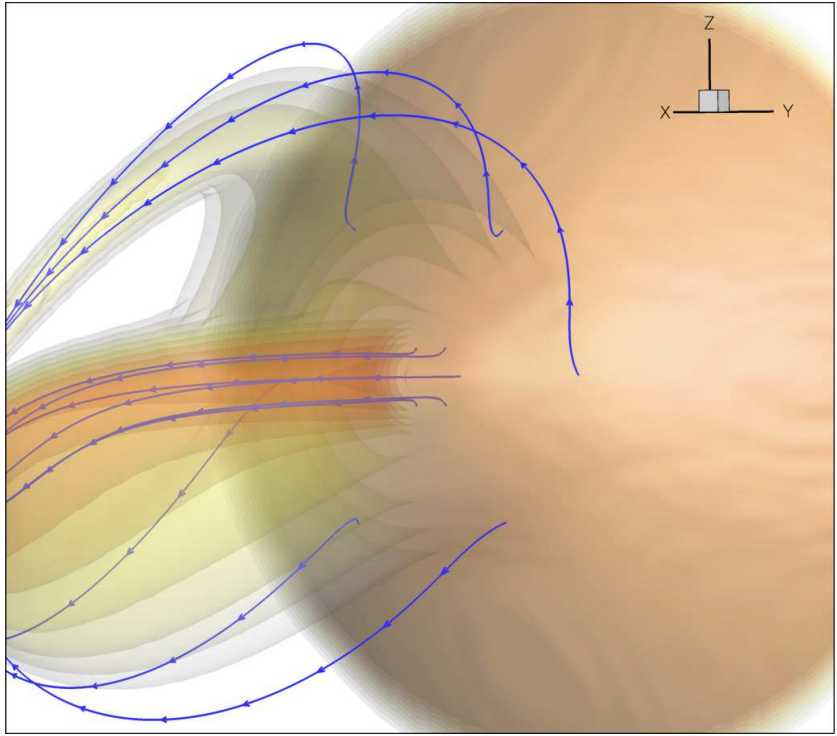}
\caption{ Flow structure in a typical polar in the region of the formation of the accretion stream. Isosurfaces of logarithm density (color) and magnetic field lines (lines with arrows) are shown.}
\label{small}
\end{figure}

The 3D structure of the flow is shown in Figs. \ref{big}
and \ref{small}. Figure \ref{big} demonstrates the full pattern of the
material flow from the donor surface. The computational domain includes part of the donor Roche lobe.
The logarithm of the density is shown in color. The
white sphere corresponds to the accretor surface. The
rotation axis of the white dwarf is directed along the
$z$ axis, marked by the blue line, and the magnetic
axis is indicated by the red line. The green lines
with arrows indicate the magnetic field lines. The
material stream from the donor moves in a wide flow
and falls onto the surface of the white dwarf at the
magnetic poles, with the main stream falling onto
the southern magnetic pole. The material first moves
along a ballistic trajectory and then slightly deviates under action of the Coriolis force. When this
material approaches the boundary of the white dwarf
magnetosphere, it becomes trapped by the magnetic field and subsequently moves along the magnetic field
lines. It is energetically more favorable for this material
to accrete onto the southern magnetic pole, since
this pole is located closer to the inner Lagrangian
point. Nevertheless, some fraction of the material forms
an accompanying flow moving toward the northern
magnetic pole.

Figure \ref{small} shows the computational results for the
flow structure near the inner Lagrangian point. The
isosurfaces of logarithm density are shown in color,
and the blue lines with arrows indicate the direction of
the flow lines. This figure demonstrates that the flow
structure has a complex character. The main stream
is formed in the immediate vicinity of the Lagrangian
point, while additional streams are formed in more
distant regions. The lower part of the accompanying stream flows into the main flow after some time.
The upper part of the accompanying stream becomes
trapped by the magnetic field of the white dwarf, and
forms a separate collimated accretion stream. An
analysis of the computational data shows that the
mass flow in this stream is about a factor of $500$ lower
than in the main stream. 

\begin{figure}[ht!]  
\centering
\begin{tabular}{cc}
\hbox{\includegraphics[width=0.45\textwidth]{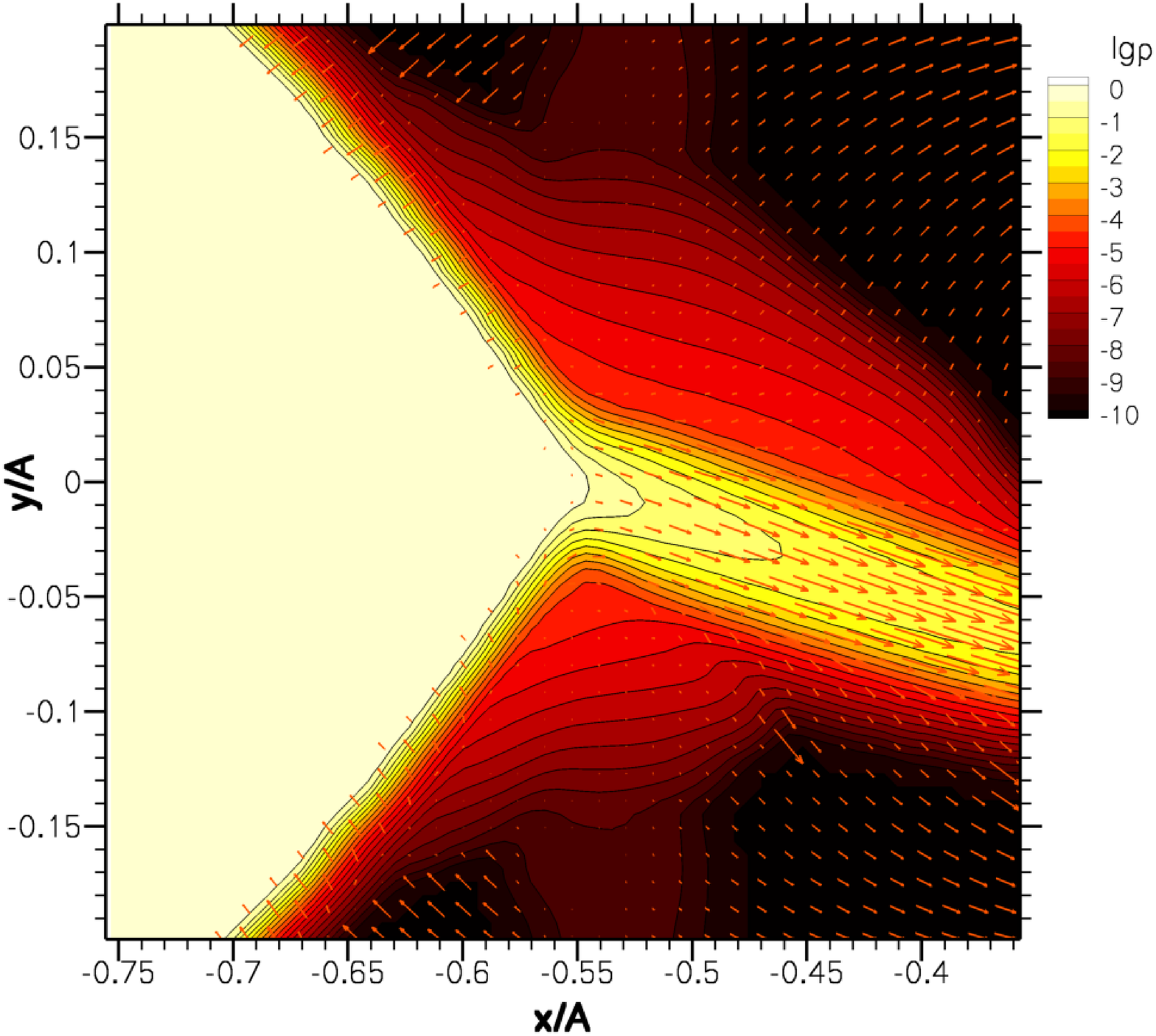}} &
\hbox{\includegraphics[width=0.45\textwidth]{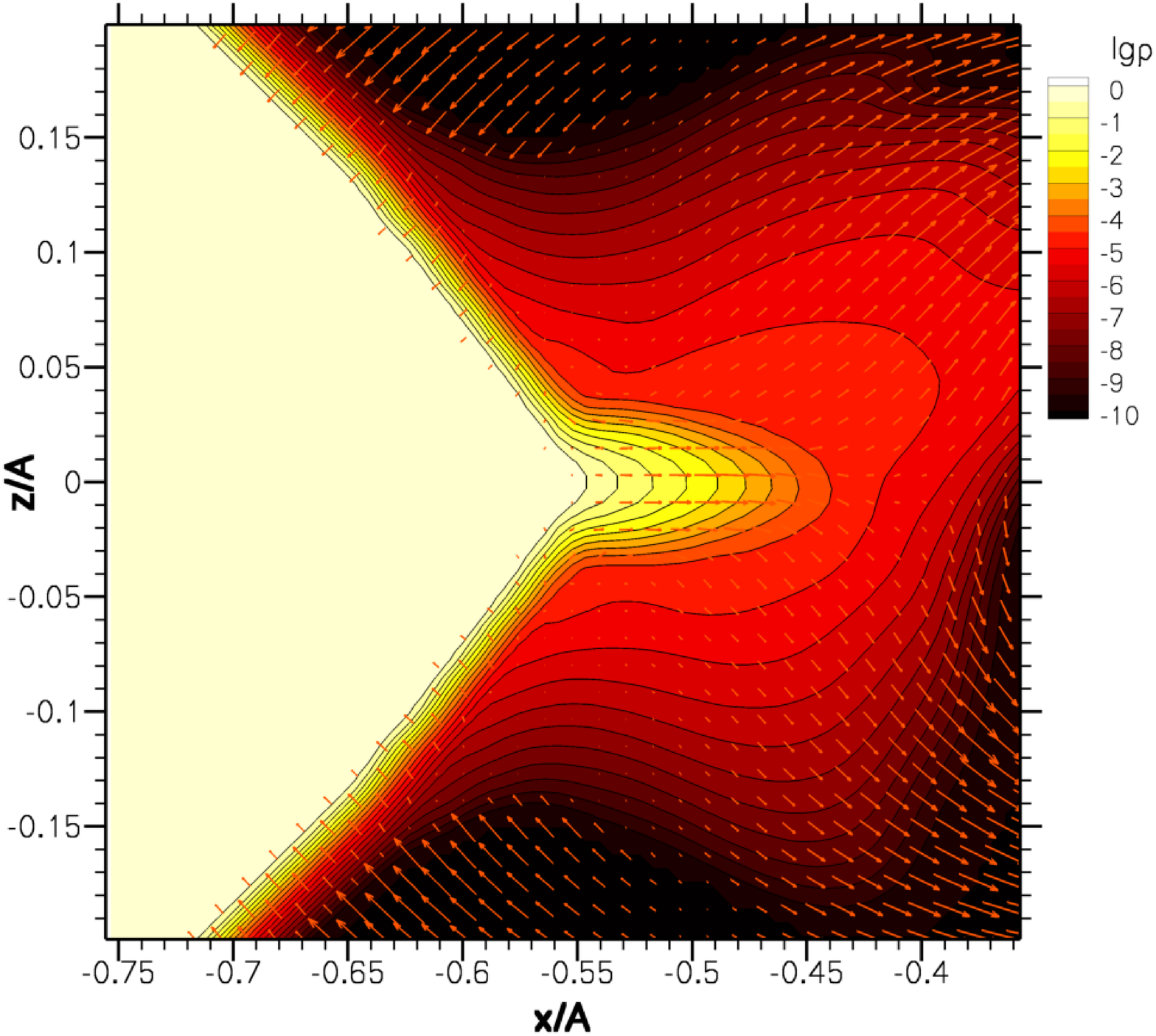}}  \\
a) $z=0$ & b) $y=0$ \\
\hbox{\includegraphics[width=0.45\textwidth]{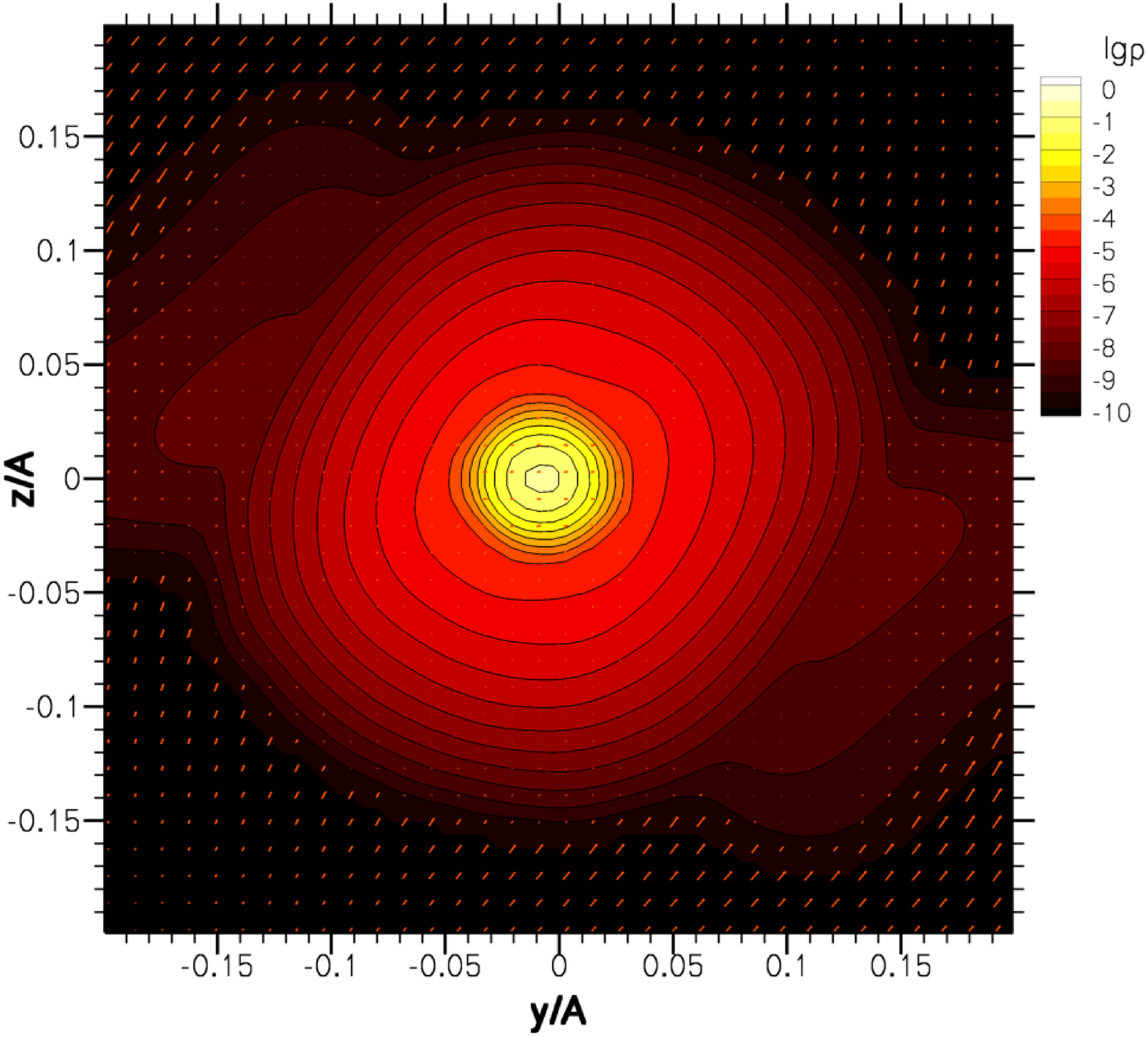}} &
\hbox{\includegraphics[width=0.45\textwidth]{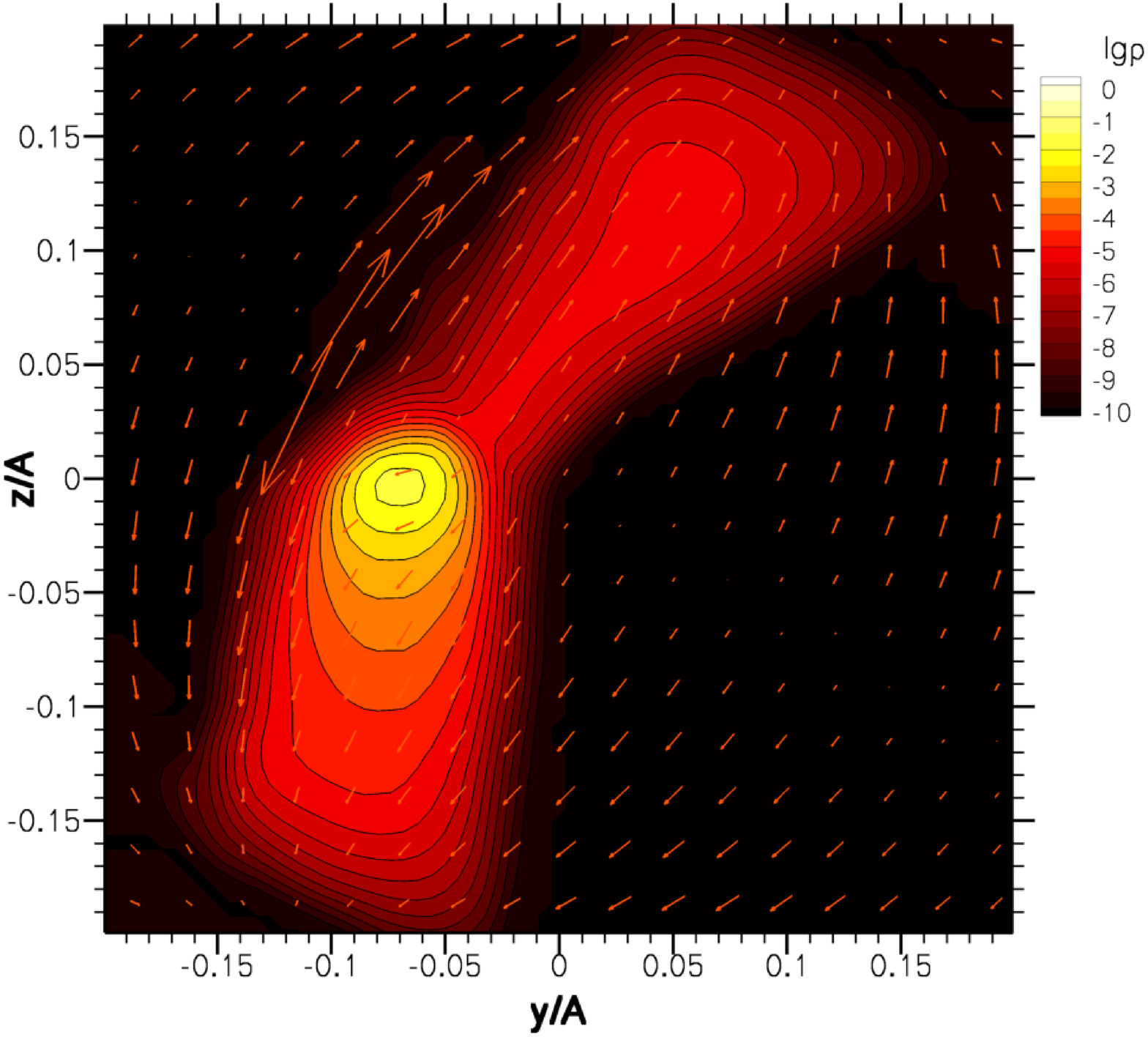}} \\
c) $x=-0.5385~A$ & d) $x=-0.3558~A$ \\
\end{tabular}
\caption{ The distribution of the density (color) and velocity (arrows) in the $xy$ plane (upper left), $xz$ plane (upper right), $yz$ plane at $x = -0.5385~A$ (lower left), and $yz$ plane at $x = -0.3558~A$ (lower right).}
\label{small2}
\end{figure}

Figure \ref{small2} demonstrates the flow structure in the
region of formation of the stream separately in three
planes: in equatorial $xy$ (upper left) and vertical
$xz$ (upper right) planes, and in the $yz$ plane where
$x=-0.5385~A$ (lower left) and $x=-0.3558~A$ (lower
right). The distribution of the density is indicated in
color, and the distribution of the velocity tangent to a
given plane is marked by arrows. Analysis of the flow
structure in the equatorial plane (upper left) shows
that most of the material from the donor moves into
the Roche lobe of the white dwarf from the vicinity
of the inner Lagrangian point. The material of the
accompanying streams located lower and higher L$_1$
has a considerably lower density. In the $xz$ plane
(upper right), the bulk of the material flows downward,
while a small portion continues to move straight
forward, forming an additional stream. In the $yz$ plane
(lower), all the material near the donor (left) moves as
one flow, but, little farther from the inner Lagrangian
point (right), the main stream is condensed and runs
downward, while a small portion of the material forms
another stream that moves upward.

\begin{figure}[ht!]  
\centering
\begin{tabular}{cc}
\hbox{\includegraphics[width=0.45\textwidth]{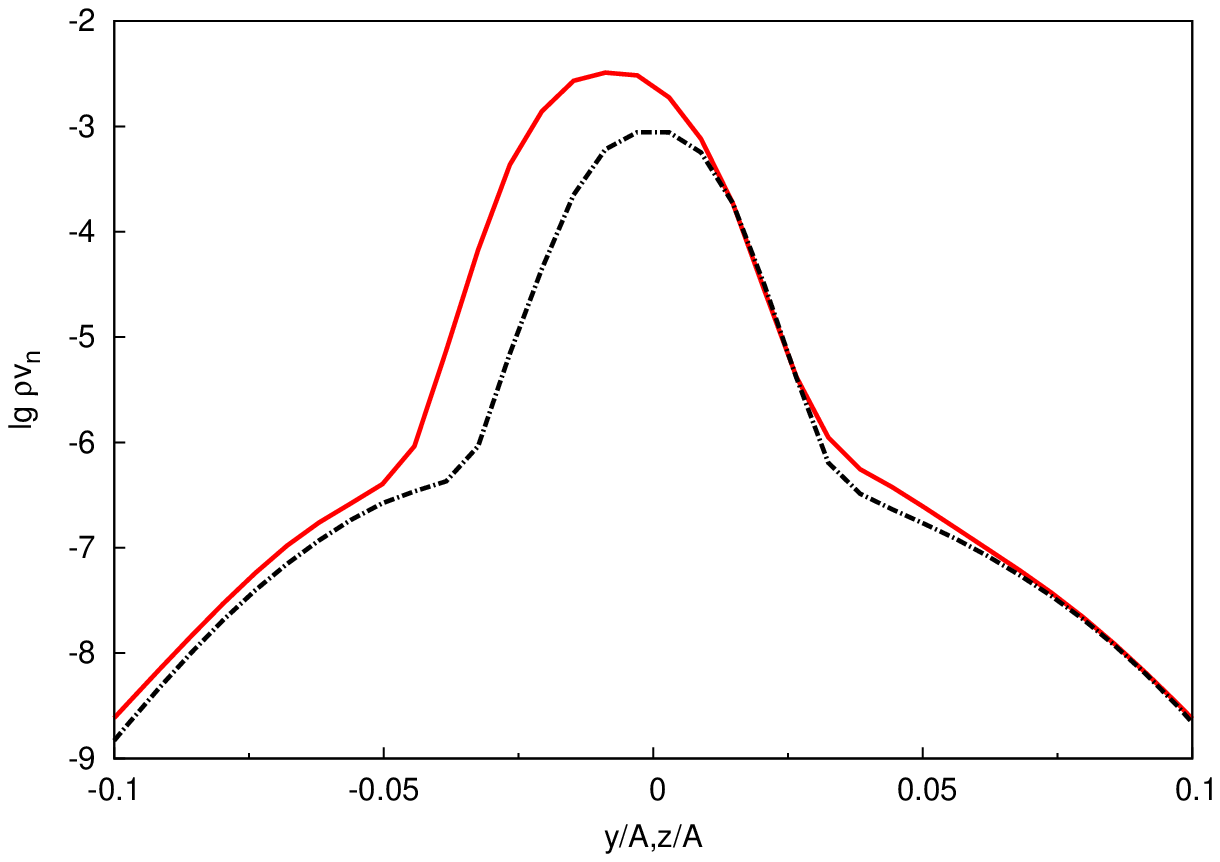}} &
\hbox{\includegraphics[width=0.45\textwidth]{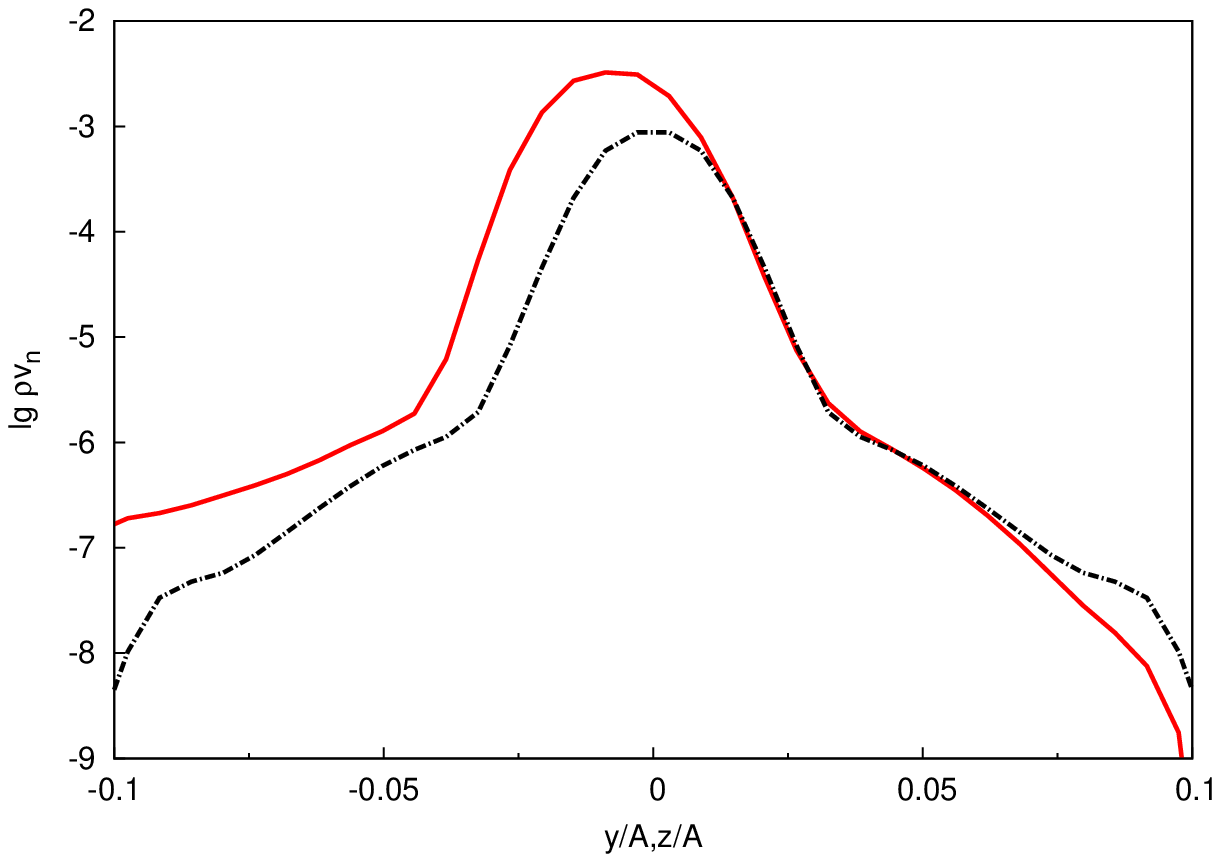}} \\
a) $B_\text{a}=10^8~\text{G}$ & b) $B_\text{a}=10^{-8}~\text{G}$ \\
\end{tabular}
\caption{ Left: logarithm of the mass flow density as a function of the cross section of the accretion flow along the $y$ (red solid curve) and $z$ (black dashed curve) is for a polar with a magnetic field of $B_\text{a}=10^8~\text{G}$ at the accretor surface. Right: same for the gas dynamical case with the same parameters, but with the surface magnetic field $B_\text{a}=10^{-8}~\text{G}$.}
\label{flow}
\end{figure}

Figure \ref{flow} shows profiles of the mass flow density $\rho v_n$  in cross sections of the accretion flow. The left plot presents profiles for the MHD case we have analyzed  ($B_\text{a} = 10^{8}~\text{G}$)); for comparison, the right plot
shows analogous profiles for the gas dynamical case
(very weak filed $B_\text{a} = 10^{-8}~\text{G}$). 
 The red and black-
dashed solid curves indicate the distributions along
the y and z coordinates, respectively. These plots
show a central flow and ''wings'' corresponding to
the accompanying stream. The distribution along the
y coordinate (red curve) is offset towards negative
values, and has a higher maximum than the profile
along the $z$ coordinate. This is due to the Coriolis
force (as is clearly visible in the upper left panel of
Fig. \ref{small2}). An analysis of the computational data shows
that the total mass flow in the main stream is almost
two orders of magnitude greater than that in the
accompanying streams. Furthermore, the total mass
flow for the gas dynamical case is approximately 10%
greater than in the MHD case, due to the influence
of the magnetic field on the formation of the stream,
which leads to suppression of the mass transfer rate.

\begin{figure}[ht!]  
\centering
\begin{tabular}{cc}
\hbox{\includegraphics[width=0.45\textwidth]{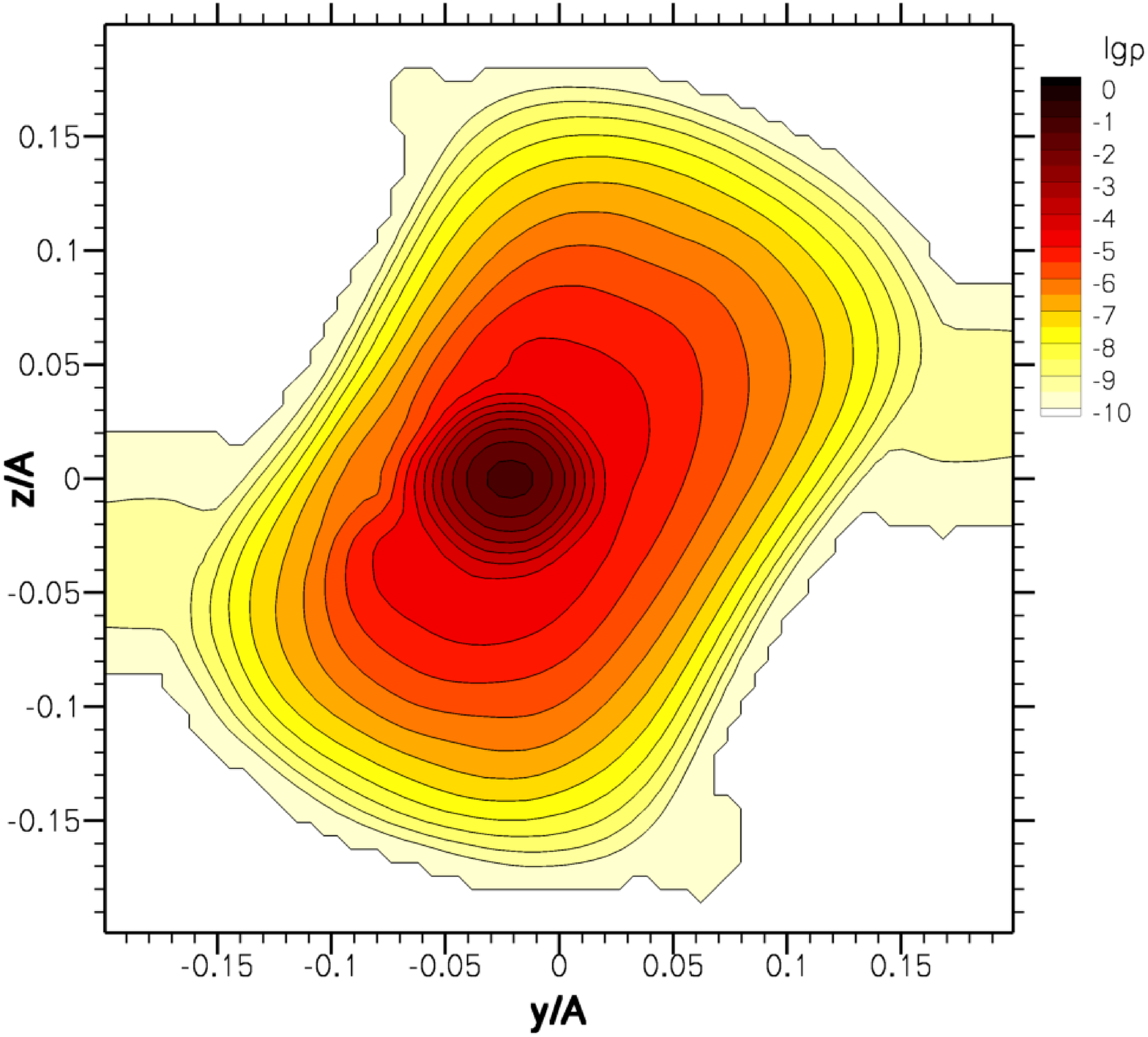}} &
\hbox{\includegraphics[width=0.45\textwidth]{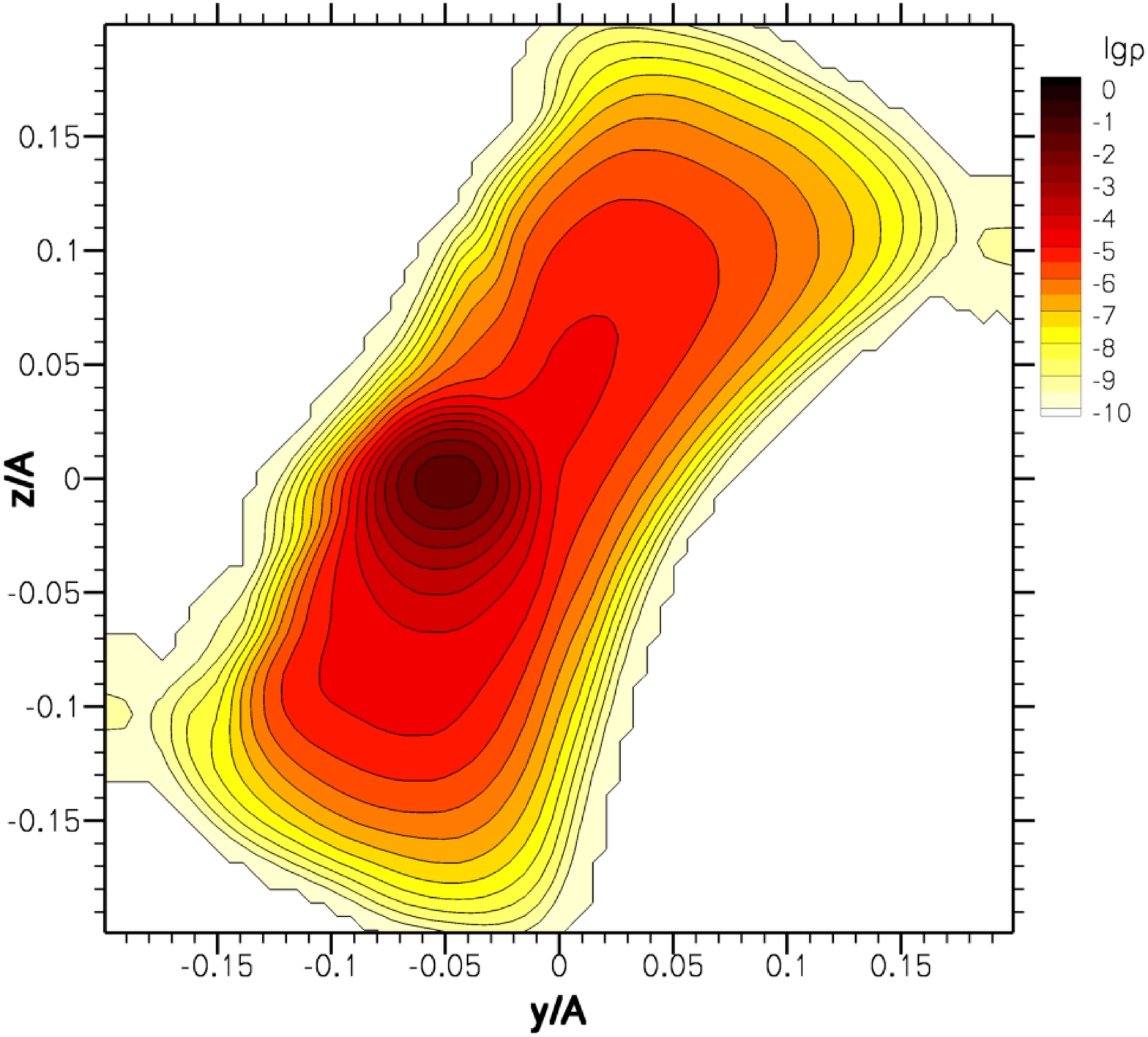}}  \\
a) $x=-0.50~A$ & b) $x=-0.42~A$ \\
\hbox{\includegraphics[width=0.45\textwidth]{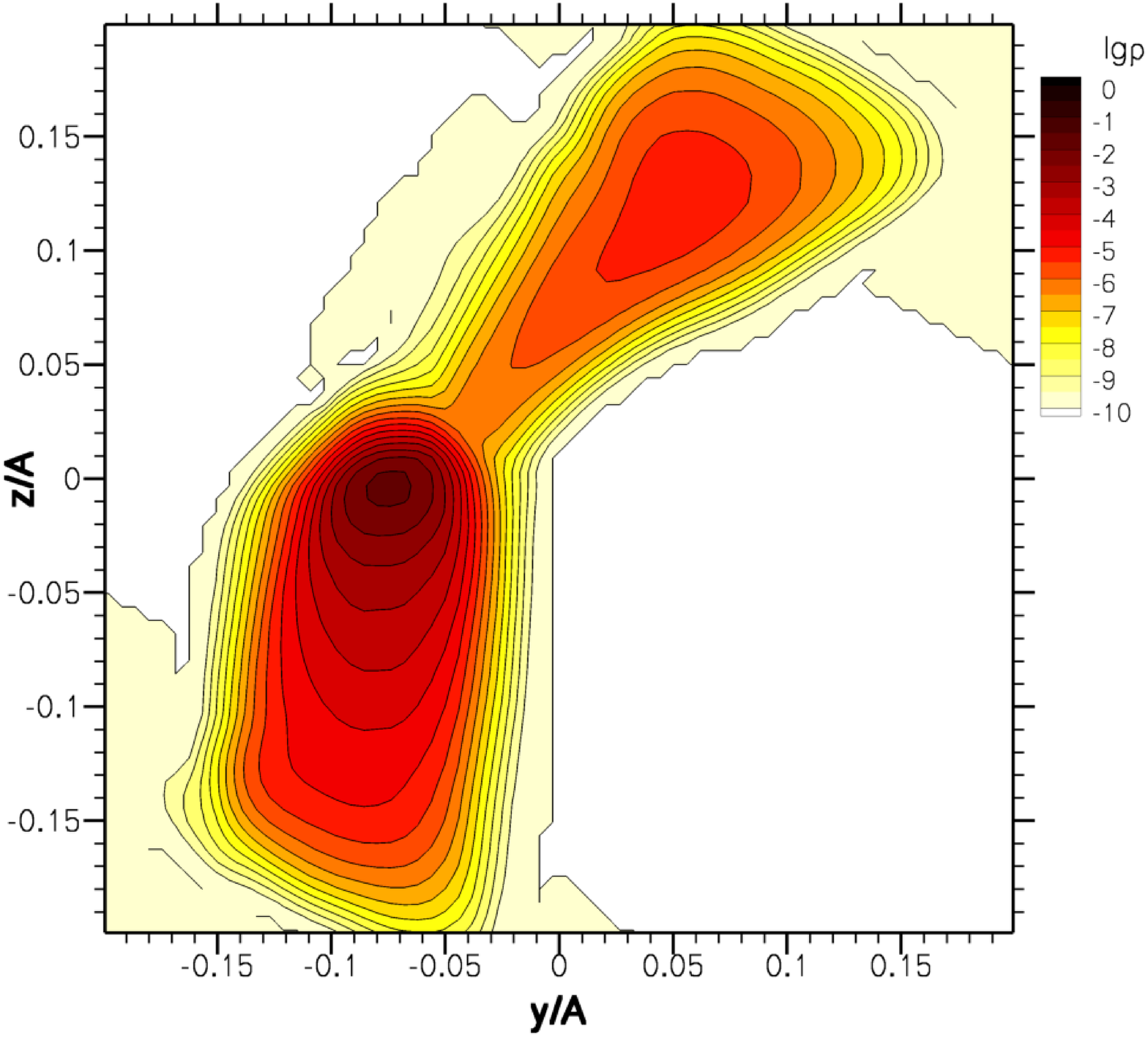}} &
\hbox{\includegraphics[width=0.45\textwidth]{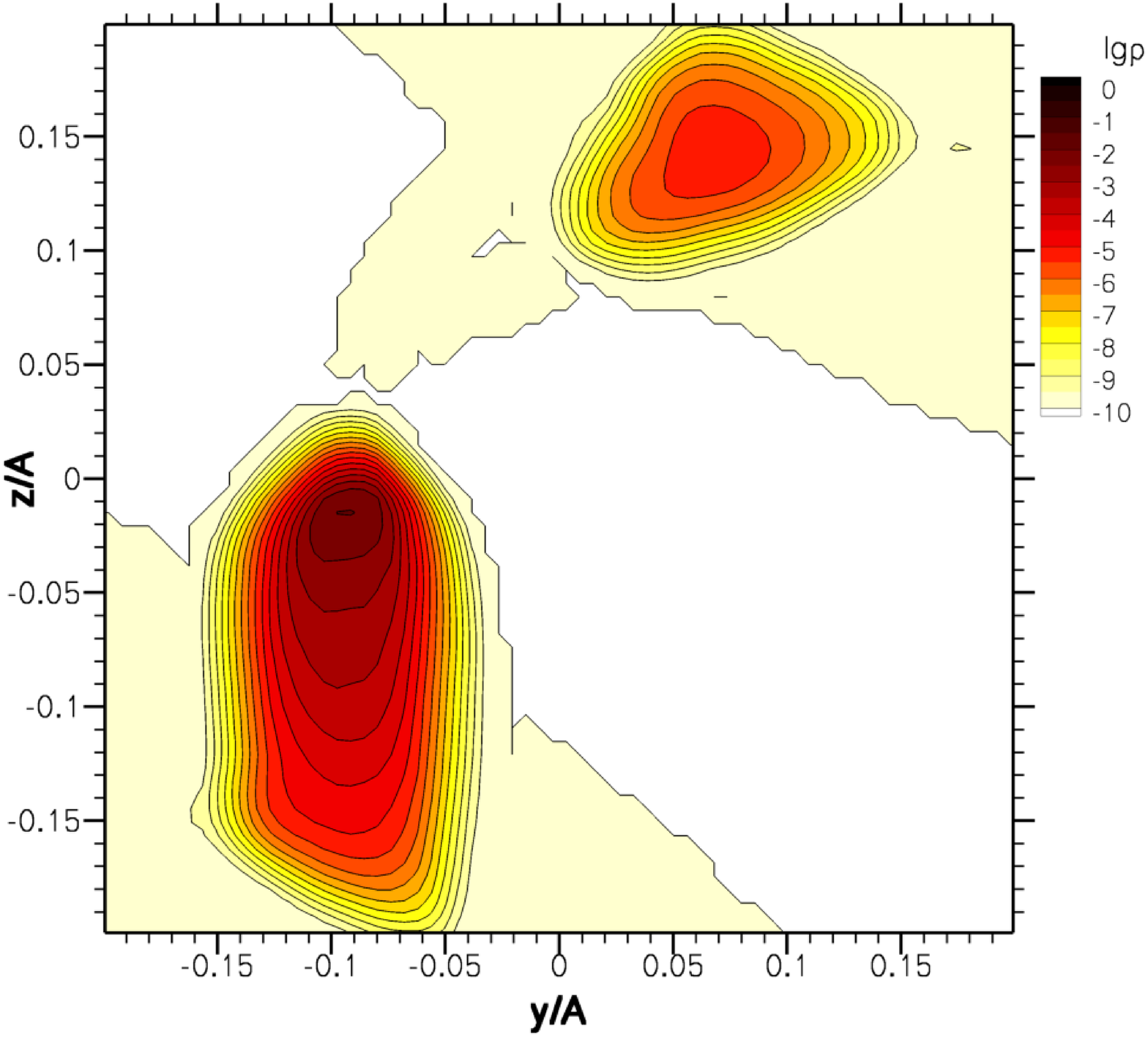}} \\
c) $x=-0.34~A$ & d) $x=-0.26~A$ \\
\hbox{\includegraphics[width=0.45\textwidth]{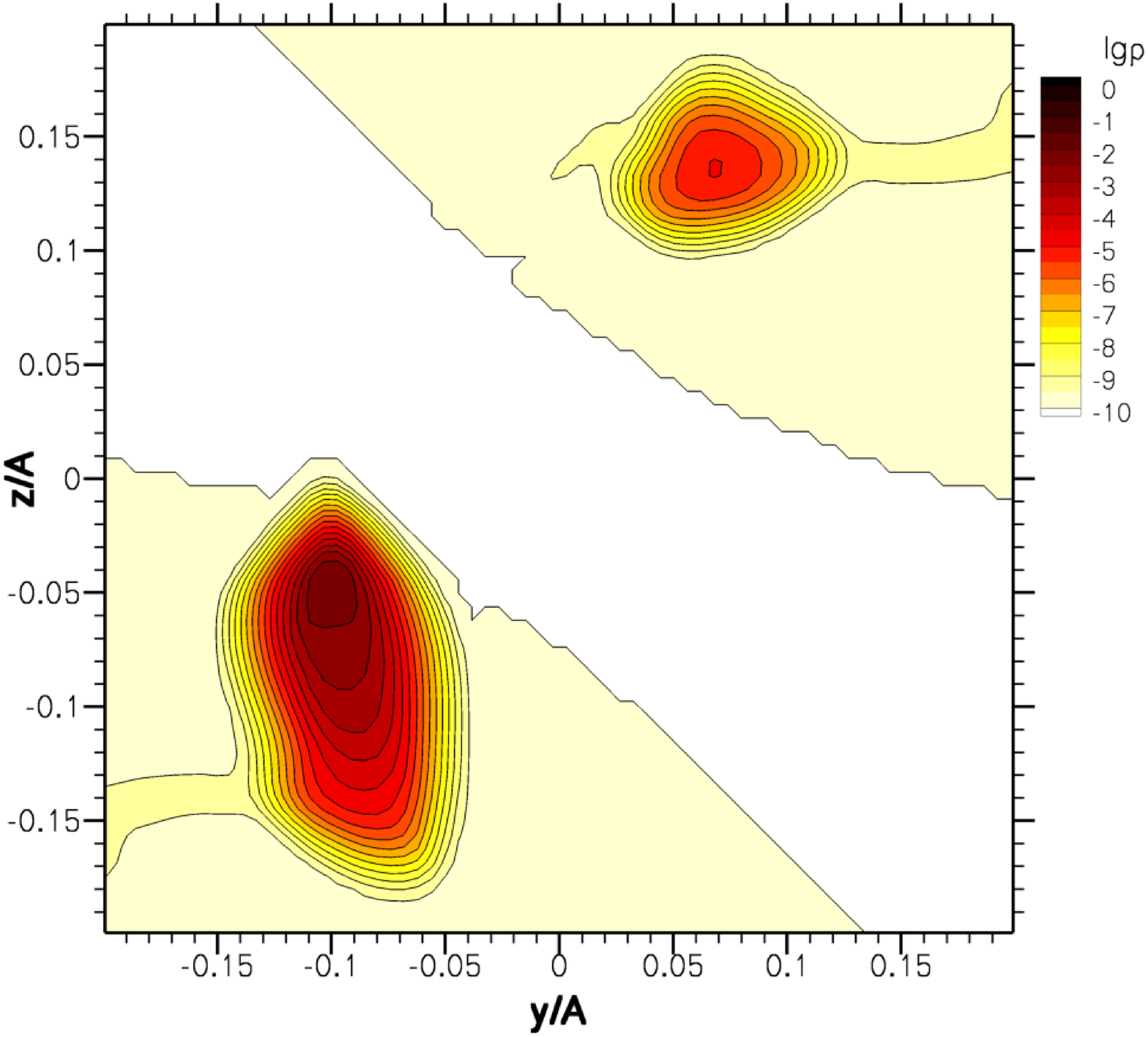}} &
\hbox{\includegraphics[width=0.45\textwidth]{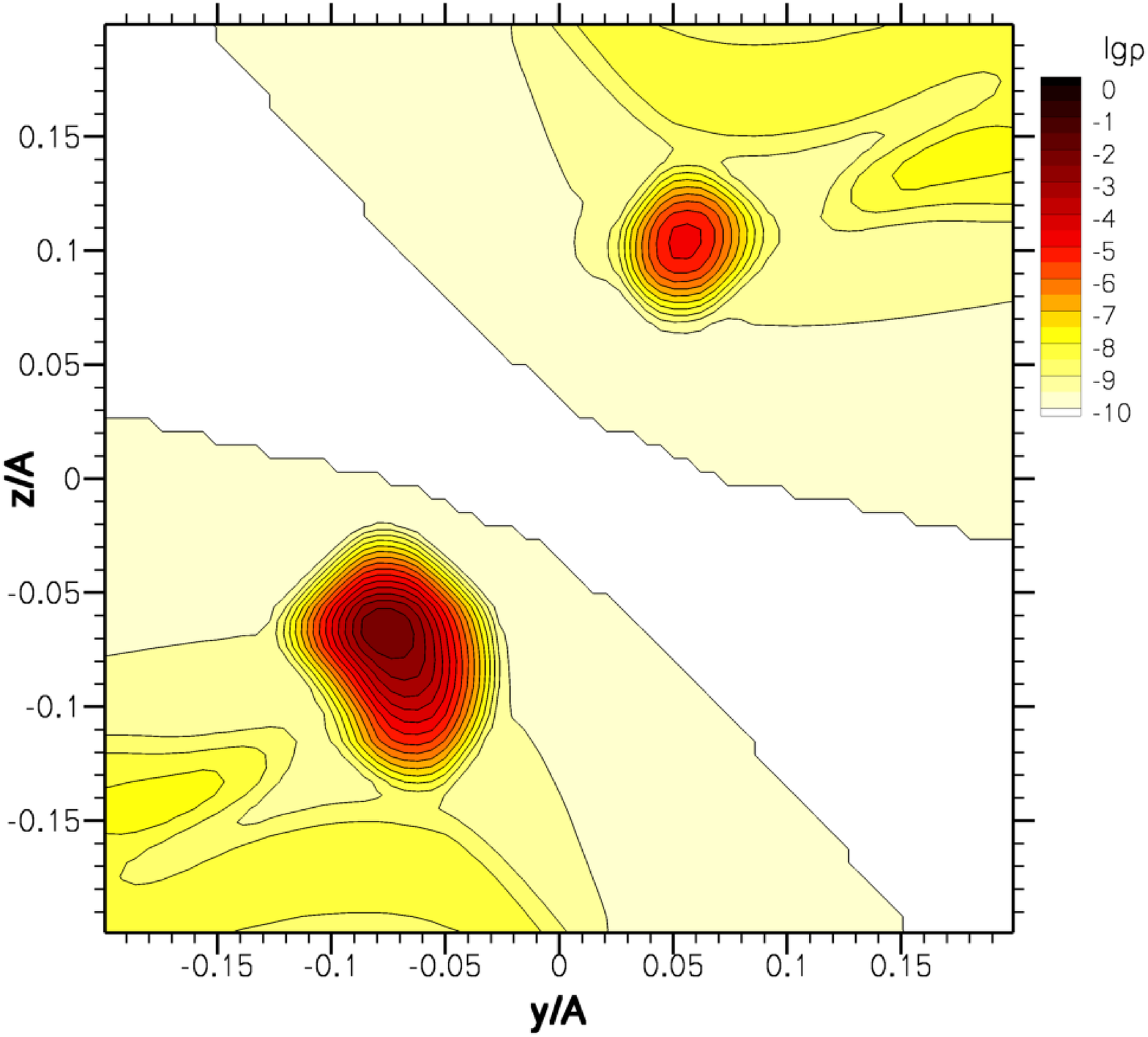}} \\
e) $x=-0.18~A$ & f) $x=-0.10~A$ \\
\end{tabular}
\caption{Distribution of the density (color) in $yz$ planes corresponding to different fixed values of x. Upper: $x = -0.50~A$ (left) and $x = -0.42~A$ (right); center: $x = -0.34~A$ (left) and $x = -0.26~A$ (right); lower: $x = -0.18~A$ (left) and $x = -0.10~A$ (right). Cross sections indicating the process of splitting of the stream are presented.}
\label{spot}
\end{figure}

Figure \ref{spot} presents density distributions (color) in
various cross sections of the $yz$ plane corresponding
to fixed x values. The x coordinate is varied in steps
of 0.08A. Recall that the coordinate of the inner
Lagrangian point is $x = -0.556~A$. The plot for $x =
-0.50~A$ (upper left) presents a section in the $yz$ plane
in the immediate vicinity of the inner Lagrangian
point. The accretion stream is extended along the
axis of symmetry of the magnetic field. Its density
increases towards the center, which is offset to the
left along the y coordinate due to the Coriolis force.
In the diagram for $x = -0.42~A$ (upper right), the
accretion flow is even more extended. The regions of
higher density are even more shifted to the left and are
somewhat extended downward, while the less dense
material are shifted upward. In the diagram for $x =
-0.34~A$ (center left panel), we can clear see two dense
centers: the lower one with higher density is shifted to
the left and downward, while the upper one with lower
density is shifted upward. A relatively thin bridge is
seen between these centers. However, this bridge has
disappeared in the diagram for $x = -0.26~A$ (center
right panel). Here, the main lower flow with higher
density and its accompanying upper flow with a lower
density are gradually moving away from each other.
In the diagram for $x = -0.18~A$ (lower left), the flows
become collimated and their cross section narrows
appreciably. Finally, in the diagram for $x = -0.10~A$
(lower right), we can see two separate collimated
accretion flows. The lower flow corresponds to material accretion onto the southern magnetic pole of the
white dwarf, and the upper flow accretion onto its
northern magnetic pole.

Analysis of these diagrams leads us to conclude
that the stream has a nonuniform cross section
structure. The magnetic field influences less dense
parts of the flow at longer distances from the white
dwarf, and more dense parts of the flow at smaller
distances. At $x = -0.34~A$, the accretion flow is
beginning to split into two separate streams. This
distance corresponds to the boundary of the magnetosphere for the peripheral, less dense part of the
accretion stream (the ''wings'' of the density profile in
Fig. \ref{flow}). The higher density part of the flow reaches
its magnetosphere boundary after some time, and,
according to this picture, should also separate onto
two distinct streams due to the action of the magnetic
field. However, the insufficient resolution of the grid
in our computations apparently hinders our detection
of this effect. In addition, these streams should again
merge as they approach the southern magnetic pole.

\begin{figure}[ht!]  
\centering
\includegraphics[width=0.9\textwidth]{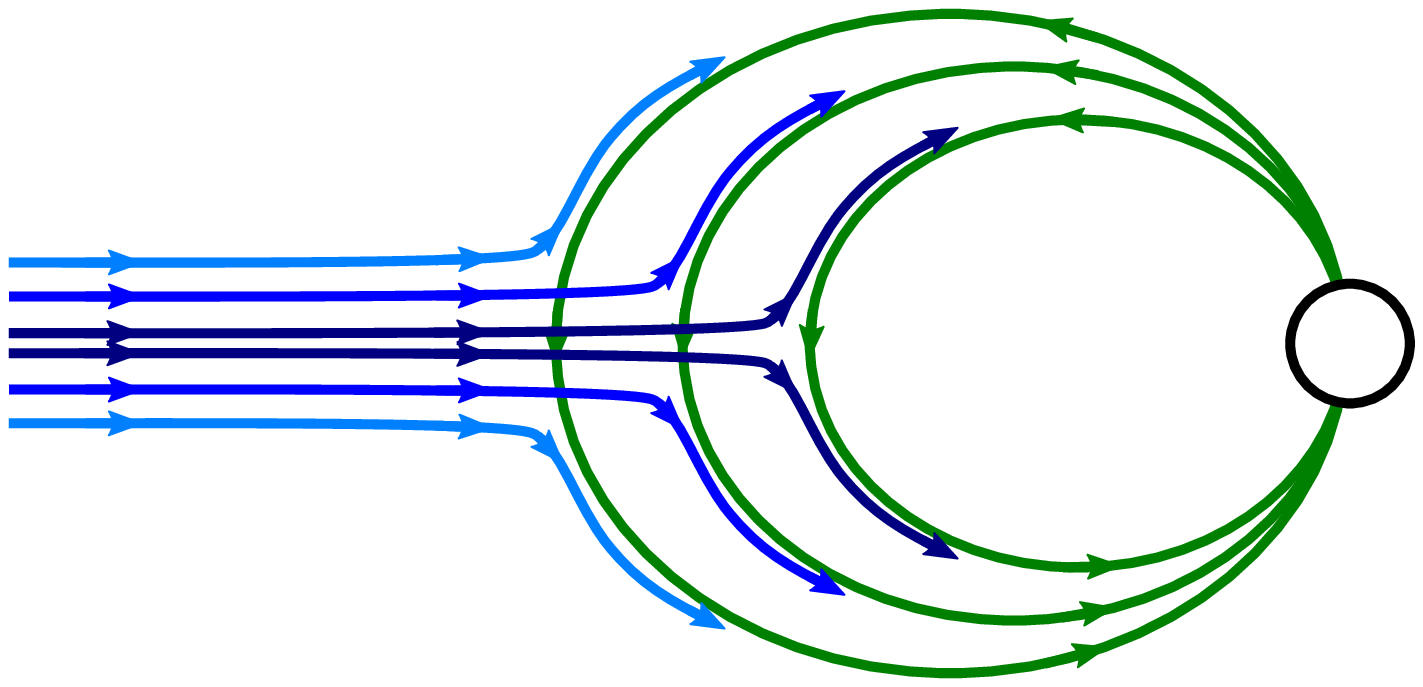}
\caption{Schematic picture of the formation of a hierarchial magnetosphere. The circle corresponds to the white dwarf, magnetic field lines are shown by the green lines with arrows, and the blue lines of various shades indicate material flow lines, with darker shades corresponding to higher density.}
\label{schema}
\end{figure}

Such a picture for the formation of the magnetosphere can be referred to as hierarchial. Above,
we described the formation of only two hierarchial
levels. However, it is clear that, in general, the
number of such levels could be infinite, since the flow
profile is continuous. Figure \ref{schema} shows a schematic of
the formation of such a hierarchial magnetosphere.
Here, the white dwarf is denoted by a circle, and its
magnetic field lines are shown by the green lines with
arrows. The material flow lines are indicated by the
blue lines with arrows. For simplicity, we present the
case for a three-layer hierarchy. Darker shades of the
flow lines correspond to higher material densities. The
peripheral parts of the flow (lightest shade of blue)
are deflected earliest by the magnetic field and form
the outer parts of the magnetosphere. More inner
and denser parts of the flow (medium-dark blue) are
deflected by the magnetic field somewhat later. The
innermost and densest parts of the flow (darkest blue)
penetrate more deeply and form the inner parts of the
magnetosphere. However, all these separate flows
can gradually merge near the magnetic poles into
unified accretion columns or curtains. This picture for
the formation of the magnetospheres in polars differs
appreciably from the classical scenario, and can be
used for more detailed analyses and interpretation of
observations.

\section{Conclusions}

We have used our 3D MHD numerical model to
investigate the flow structure in polars. This model
is based on a modified MHD approximation, which
describes the plasma dynamics in the case of strong
external magnetic fields, taking into account Alfv{\`e}n
wave turbulence in the presence of low magnetic
Reynolds numbers. The computational domain fully
encompassed the Roche lobe of the accretor, as well
as part of the Roche lobe of the donor. This has
enabled us to describe the formation of the outflow
from the donor envelope in the vicinity of the inner
Lagrangian point by a natural way using our model.
Furthermore, in contrast to our preceding studies, we
have used the energy equation rather than the entropy
equation in this model.

We have carried out 3D MHD numerical simulations of mass transfer from the donor into the Roche
lobe of the accretor. As an example, we have considered a polar with a magnetic field at the white dwarf
surface of $B_\text{a} = 10^8~\text{G}$, whose parameters correspond to SS Cyg. The structure of the formed flow
was studied, paying special attention to the vicinity of
the inner Lagrangian point, where the accretion flow
is formed.
The results of numerical simulations show that
the material flowing from the donor’s envelope into the
Roche lobe of the accretor forms collimated accretion flows that move toward the magnetic poles of
the white dwarf. The flow formed in the vicinity of
the inner Lagrangian point splits onto two separated
streams due to the magnetic field. The bulk of the
material is deflected downwards by the Coriolis force,
and moves toward the southern magnetic pole. The
less dense flow is coupled with the magnetic field lines
of the accretor earlier, and drops onto the white dwarf
surface in the region of its northern magnetic pole.
The interaction of the accretion flow material from
the donor envelope with the magnetic field leads to
the formation of a hierarchial structure for the magnetosphere. The less dense (peripheral) parts of the
flow are influenced by the magnetic field at longer
distances from the accretor, and form the outer magnetosphere. More inner and denser parts of the flow
are deflected by the magnetic field at closer distance
to the accretor. The innermost and densest parts of
the flow penetrate through the magnetic field and form
the innermost parts of the magnetosphere. However,
all these separate streams should merge near the
magnetic poles and form the accretion columns or
curtains at the surface of the white dwarf \cite{Semena:2012}. This
scenario for the formation of hierarchial magnetospheres in polars differs appreciably from the classical picture; taking this into account can affect the results of analyses and the interpretation of observations.

\section*{Acknowledgments}
P.B.I. was supported by the Russian Foundation for Basic Research (project 16-32-00909). This work
has been carried out using computing resources of the federal collective usage center Complex for Simulation and Data Processing for Mega-science Facilities at NRC ''Kurchatov Institute'', http://ckp.nrcki.ru/.

\end{document}